\documentclass[twocolumn,showpacs,preprintnumbers,amsmath,amssymb,showkeys]{revtex4}
\usepackage{graphicx}
\usepackage{dcolumn}
\usepackage{bm}

\begin{document}

\title{Effective numbers of charge carriers in doped graphene: The generalized Fermi liquid approach}
 \author{I. Kup\v{c}i\'{c}, G. Nik\v{s}i\'{c}, Z. Rukelj, and D. Pelc}     
 \address{
   Department of Physics, Faculty of Science, University of Zagreb, 
   P.O. Box 331,  HR-10002 Zagreb,  Croatia}

\begin{abstract} 
The single-band current-dipole Kubo formula for the dynamical conductivity of heavily doped graphene from 
Kup\v{c}i\'{c} [Phys. Rev. B {\bf 91}, 205428 (2015)] is extended to a two-band model for 
conduction $\pi$ electrons in lightly doped graphene.
	Using {\it a posteriori} relaxation-time approximation in the two-band quantum transport equations, 
with two different relaxation rates and one quasi-particle lifetime, we explain a seemingly inconsistent dependence 
of the dc conductivity $\sigma^{\rm dc}_{\alpha \alpha}$ of ultraclean and dirty lightly doped graphene samples on electron doping,
{{{\rm in a way consistent with}}} 
the charge continuity equation.
	It is also shown that the intraband contribution to the effective number of conduction electrons in
$\sigma^{\rm dc}_{\alpha \alpha}$ vanishes at $T=0$ K in the ultraclean regime, 
but it remains finite in the dirty regime.
	The present model is shown to be consistent with a picture in which the intraband and interband contributions 
to $\sigma^{\rm dc}_{\alpha \alpha}$ are characterized by two different mobilities of conduction electrons, 
the values of which are well below the widely accepted value of mobility in ultraclean graphene.
	The dispersions of Dirac and $\pi$ plasmon resonances are reexamined to show that the present, relatively simple
expression for the dynamical conductivity tensor can be used to study simultaneously single-particle excitations 
in the dc and optical conductivity and collective excitations in energy loss spectroscopy experiments.
\end{abstract}
%
%
%
\maketitle

\section{Introduction}
In quantum field theory, Fermi liquids are completely described in terms of single-electron Green's functions and renormalized 
charge/current vertex functions \cite{Abrikosov75,Ziman88}.
	The Green's functions satisfy the corresponding Dyson equations, and the vertex functions the Bethe-Salpeter equations.
	In electronic systems with parabolic dispersions and weak residual interactions 
{{{\rm the problem of solving}}}
these two self-consistent equations
{{{\rm reduces to analyzing}}}
the semiclassical Landau-Silin transport equations for nonequilibrium distribution functions \cite{Pines89,Platzman73}.
	The details about electron-electron 
interactions are hidden in renormalized electron dispersions and in transport relaxation rates.
	Electrodynamic properties of such systems are well known and are usually expressed in terms of the nominal 
concentration of conduction electrons, or the electron density of states at the Fermi level, 
and in terms of the well-known Landau scattering functions.
	The effective mass of conduction electrons and the effective density of states are 
introduced to describe the effects of residual electron-electron interactions in the simplest way.
	Almost all observables look the same as in the theory of noninteracting fermions, 
with the exception that the electron mass and the electron density of states are replaced by their effective values.

In contrast, when the residual interactions among conduction electrons are strong, e.g. in underdoped cuprates \cite{Schrieffer07}, 
there is no way to simplify the original self-consistent equations.
	In addition, when these equations are treated beyond the leading (Hartree-Fock) approximation,
it is necessary to replace bare electron-electron interactions by irreducible four-point interactions.
	Consequently, the resulting expressions for different transport coefficients, for 
{{{\rm the real and imaginary parts of}}} 
the dielectric function, and for many other response functions
cannot be mapped onto standard Fermi liquid expressions.
        The concentration of conduction electrons and the bare density of states 
are no longer quantities which enter in observables as multiplicative parameters.
        They are replaced by different forms of the effective number of charge carriers 
and different effective densities of states \cite{Kupcic07}.
	More importantly, in such a general formulation of the response functions, 
there is no need for using concepts such as the transport or optical electron mass.
     
The residual electron-electron interactions in graphene are presumably weak, but the electron dispersions 
are very different from the parabolic dispersion.
	Therefore, to understand electrodynamic properties of 
pristine and doped graphene, as well as to answer open questions regarding the behaviour of conduction $\pi$ electrons 
{{{\rm in the presence of external electromagnetic fields,}}}
we are forced once again to use the original Dyson and Bethe-Salpeter equations
instead of the semiclassical Landau-Silin equations, and to treat the dispersions 
of $\pi$ electrons beyond the Dirac cone approximation.
	However, the leading approximation for irreducible four-point interactions can still be used \cite{Kupcic15}.

The paper is organized as follows.
	In Sec.~II, we consider the multiband quantum transport equations in the Hartree-Fock approximation \cite{Kupcic13}. 
	These equations are expected to be appropriate for studying relaxation processes in graphene at low enough temperatures
where the conduction electrons are scattered primarily by static disorder and by phonons.
	In these equations there are two types of damping energies: the single-electron damping energy 
[i.e., the half width of the quasi-particle peak in angle-resolve photoemission spectra 
(ARPES) \cite{Bostwick07,Pletikosic12}] and the electron-hole damping energies 
(i.e., the intraband and interband relaxation rates in the dc and optical conductivity \cite{Novoselov05,Bolotin08,Li08}). 
	The charge continuity equation is responsible for the fact that ${\bf q} \approx {\bf 0}$ scattering processes 
{{{\rm drop out of}}} 
the intraband electron-hole damping energies.
	On the contrary, the single-electron damping energies 
{{{\rm depend quite drastically on}}}
the intensity of these scattering processes.
	We use the multiband Ward identity relations \cite{Kupcic15,Schrieffer64}
to determine the structure of the effective number of charge carriers 
in the two-band version of the transverse conductivity sum rule, and show that this effective number does not depend on the 
relaxation rates.
	In Sec.~III, low-order perturbation theory is used to emphasize different roles played 
by vertex corrections in the intraband and interband quantum transport equations.
	In Sec.~IV, 
{{{\rm the observed dc conductivity}}} 
of lightly doped graphene samples \cite{Bolotin08} is analyzed 
by using the current-dipole conductivity formula, with particular care devoted to two types of damping energies 
and to two types of contributions to the effective number of charge carriers.
	We use the relaxation-time approximation, for simplicity, with reasonable values of the relaxation rates.
	They agree with both experimental observation \cite{Bolotin08}
and with theoretical predictions 
for the dependence of the single-electron damping energy on the Fermi energy $E_{\rm F}$ from Refs.~\cite{Ando98,Ando02,Peres06}. 
	In Sec.~V, we write the current-dipole conductivity formula 
{{{\rm in the alternative form}}} 
and briefly discuss disadvantages of this conductivity formula with respect to the current-dipole formula from Sec.~IV.
	In Sec.~VI, the dispersions of Dirac and $\pi$ plasmons are reexamined to emphasize that the present current-dipole 
approach can be used to study the dc and dynamical conductivity 
{{{\rm on an equal footing with}}} 
finite ${\bf q}$ properties of doped graphene.
	Section VII contains concluding remarks.

\section{Transverse conductivity sum rule}

\begin{figure}
   \centerline{\includegraphics[width=20pc]{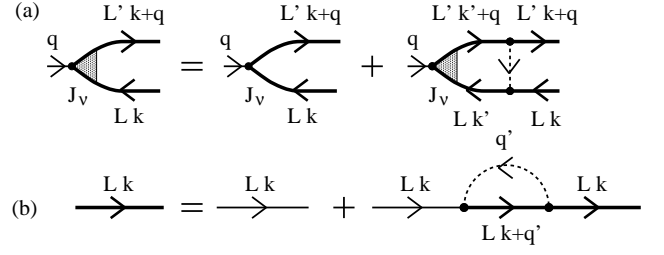}}
   \caption{(a) The Bethe-Salpeter equations for the auxiliary electron-hole propagators
   $\Phi_{\nu}^{LL'} ({\bf k},{\bf k}_+,  {\it i}  \omega_n, {\it i}  \omega_{n+})$
   in the Hartree-Fock approximation.
   The dashed line represents the force-force correlation function ${\cal F}_\lambda ({\bf k}'-{\bf k}, {\it i} \nu_{m})$
   and the bold solid lines are the single-electron propagators ${\cal G}_{L} ({\bf k}, {\it i} \omega_n)$.
   (b) The Dyson equation for ${\cal G}_{L} ({\bf k}, {\it i} \omega_n)$ in the same approximation.
   }
  \end{figure}	
  
{{{\rm An appropriate starting point for}}}
the microscopic examination of the relaxation processes in weakly interacting multiband electronic systems in which 
local field effects are absent (the two-band model for $\pi$ electrons in doped graphene being an example)
are the Bethe-Salpeter equations for the electron-hole propagators 
$\Phi^{LL'}_{\nu} ({\bf k},{\bf k}_+, {\it i} \omega_n, {\it i} \omega_{n+} ) $ 
shown in the Hartree-Fock  approximation, 
\begin{eqnarray}
&& \hspace{-5mm}
\Phi^{LL'}_{\nu} ({\bf k},{\bf k}_+, {\it i} \omega_n, {\it i} \omega_{n+} ) 
= \frac{1}{\hbar^2} {\cal G}_{L'} ({\bf k}_+, {\it i} \omega_{n+} ) {\cal G}_{L} ({\bf k}, {\it i} \omega_n ) 
\nonumber \\
&&  \hspace{12mm} \times 
\bigg\{ J_{\nu}^{L'L}({\bf k}_+,{\bf k}) - 
\sum_{\lambda {\bf k}'} \frac{1}{\beta}  \sum_{{\it i} \omega_m} {\cal F}_\lambda ({\bf k}'-{\bf k},{\it i} \nu_{m})
\nonumber \\
&&  \hspace{12mm}
\times \Phi^{LL'}_{\nu} ({\bf k}',{\bf k}_+', {\it i} \omega_m, {\it i} \omega_{m+} ) \bigg\}.
\label{eq1}
\end{eqnarray}
	They are illustrated in Fig.~1(a).
	Here, $L$ is the band index, ${\bf k}_+ = {\bf k}+{\bf q}$, ${\it i}\omega_{n+}={\it i}\omega_{n}+ {\it i} \nu_n$, 
${\it i} \nu_{m} = {\it i}\omega_{m}-{\it i}\omega_{n}$, and $\nu = 0, \alpha$
($L = \pi, \pi^*$ and $\alpha = x, y$ in graphene).
	The related quantum transport equations are of the form \cite{Kupcic13,Kupcic15}
\begin{eqnarray}
&& \hspace{-5mm}
D_{LL'}^{-1} ({\bf k},{\bf k}_+, {\it i} \omega_n, {\it i} \omega_{n+} ) 
\Phi^{LL'}_{\nu} ({\bf k},{\bf k}_+, {\it i} \omega_n, {\it i} \omega_{n+} ) 
\nonumber \\
&&  \hspace{10mm}
= \frac{1}{\hbar^2} \big[ {\cal G}_{L} ({\bf k}, {\it i} \omega_n ) - {\cal G}_{L'} ({\bf k}_+, {\it i} \omega_{n+} ) \big] 
\nonumber \\
&&  \hspace{12mm} \times 
\bigg\{ J_{\nu}^{L'L}({\bf k}_+,{\bf k}) - 
\sum_{\lambda {\bf k}'} \frac{1}{\beta}  \sum_{{\it i} \omega_m} {\cal F}_\lambda ({\bf k}'-{\bf k},{\it i} \nu_{m})
\nonumber \\
&&  \hspace{12mm}
\times \Phi^{LL'}_{\nu} ({\bf k}',{\bf k}_+', {\it i} \omega_m, {\it i} \omega_{m+} ) \bigg\}.
\label{eq2}
\end{eqnarray}
	In these two equations, 
\begin{eqnarray}
&& \hspace{-5mm}
\hbar^2 \Phi_{\nu}^{LL'} ({\bf k},{\bf k}_+,{\it i}\omega_n,{\it i}\omega_{n+})  = 
{\cal G}_{L'} ({\bf k}_+, {\it i} \omega_{n+}) {\cal G}_{L} ({\bf k}, {\it i} \omega_{n})
\nonumber \\
&&  \hspace{35mm}
\times \Gamma^{L'L}_{\nu}({\bf k}_+,{\bf k},{\it i}\omega_{n+},{\it i}\omega_n)
\label{eq3}
\end{eqnarray}
is the auxiliary RPA irreducible electron-hole propagator, 
which is the product of two single-electron Green's functions and the renormalized vertex 
$\Gamma^{L'L}_{\nu}({\bf k}_+,{\bf k},{\it i}\omega_{n+},{\it i}\omega_n)$ \cite{Abrikosov75,Dzyaloshinskii73,Schrieffer64,Kupcic15}.
	Moreover,
\begin{eqnarray}
&& \hspace{-5mm}
\hbar \Sigma_L ({\bf k}, {\it i} \omega_n ) \approx  -\sum_{\lambda {\bf k}'} \frac{1}{\beta \hbar} 
\sum_{{\it i}  \omega_m} {\cal G}_{L} ({\bf k}', {\it i} \omega_{m}) {\cal F}_\lambda ({\bf k}'-{\bf k}, {\it i} \nu_{m})
\nonumber \\
\label{eq4}
\end{eqnarray}
is the single-electron self-energy in the Dyson equation from Fig.~1(b);   
\begin{eqnarray}
&&  \hspace{-10mm}
D_{LL'}^{-1}({\bf k},{\bf k}_+, {\it i} \omega_n, {\it i} \omega_{n+}) 
=  {\it i} \nu_{n} + \varepsilon^0_{LL'}({\bf k},{\bf k}_+)/\hbar 
\nonumber \\
&& \hspace{20mm}  
+ \Sigma_L ({\bf k}, {\rm i} \omega_n ) - \Sigma_{L'} ({\bf k}_+, {\rm i} \omega_{n+})
\label{eq5}
\end{eqnarray}
is a useful abbreviation, and
${\cal F}_\lambda ({\bf k}'-{\bf k}, {\it i} \nu_m )$ is the force-force 
correlation function in the scattering channel labeled by the index $\lambda$ \cite{Mahan90,Kupcic13}.
	Scattering from static disorder and from phonons is described by the Hamiltonian 
$H_1'$ and scattering from other electrons by the nonretarded Coulomb forces in $H_2'$.
	{{{\rm With little loss of generality, we restrict the analysis to the case where}}}
the electron
{{{\rm does not change the band when it is scattered.}}}
	The generalization is straightforward, and as is shown in Refs.~\cite{Ando98,Peres06},
it must be done when considering scattering processes in pristine and lightly doped graphene 
beyond the relaxation-time approximation.
	Finally,
\begin{eqnarray}
&& \hspace{-10mm}
J_0^{LL'}({\bf k},{\bf k}_+)  
= \sum_\alpha q_\alpha \frac{ \hbar J_\alpha^{LL'} ({\bf k},{\bf k}_+)}{
\varepsilon_{L'L}^0 ({\bf k}_+,{\bf k})}
\label{eq6}
\end{eqnarray}
are the bare intraband ($L'=L$) and interband ($L' \neq L$) charge vertex functions, 
$J_\alpha^{LL'} ({\bf k},{\bf k}_+)$ is the bare current vertex \cite{Kupcic14}, 
$\varepsilon_{L}^0 ({\bf k})$ is the bare electron dispersion measured with respect to the chemical potential $\mu$, and
$\varepsilon_{L'L}^0 ({\bf k}_+,{\bf k}) =\varepsilon_{L'}^0 ({\bf k}_+)-\varepsilon_{L}^0 ({\bf k})$.
	The relation (\ref{eq6}) can be easily proven in any exactly solvable multiband model.
	It is
{{{\rm a direct consequence of}}}
the gauge invariant form of the coupling Hamiltonian in which the conduction electrons couple
to external electromagnetic fields \cite{Kupcic13,Kupcic15}.

It is generally agreed that the quantum transport equations (\ref{eq2}) are a good starting point in the longitudinal response theory
\cite{Vollhardt80,Kupcic15,Kupcic13},
while the Bethe-Salpeter equations (\ref{eq1}) are more appropriate for considering the response 
to transverse electromagnetic fields, in particular in the case where the vertex corrections 
[the second term in the curly braces in Eqs.~(\ref{eq1}) and (\ref{eq2})] are neglected \cite{Mahan90,Ando98,Ando02}.
	As pointed out in Ref.~\cite{Kupcic13}, there is a direct link between this form of the quantum transport equations 
and both the semiclassical Boltzmann transport equations and the Landau-Silin equations.
	Equations (\ref{eq1}) and (\ref{eq2}) are simplified versions of the general equations \cite{Dzyaloshinskii73,Kupcic15} in which
irreducible four-point interactions are replaced by the force-force correlation function 
${\cal F}_\lambda ({\bf k}'-{\bf k},{\it i} \nu_{m})$.
	It is well known that if we are interested in the relaxation processes associated with the electron scattering 
from other electrons, the next corrections must be included \cite{Kupcic14,Kupcic15}. 
	For the electron scattering from static disorder and from phonons, this approximation is sufficient.

\subsection{Gauge invariance of the response theory}
{{{\rm It is not hard to verify that}}}
the auxiliary electron-hole propagators $\Phi^{LL'}_{\nu} ({\bf k},{\bf k}_+, {\it i} \omega_n, {\it i} \omega_{n+})$ 
in Eqs.~(\ref{eq1}) and (\ref{eq2}) must satisfy the following equations:
\begin{eqnarray}
&& \hspace{-10mm}
\sum_{LL'} \frac{1}{V} \sum_{{\bf k} \sigma} 
\big[\omega J_0^{LL'} ({\bf k},{\bf k}_+) - 
\sum_\alpha q_\alpha J_\alpha^{LL'} ({\bf k},{\bf k}_+)\big] 
\nonumber \\
&& \hspace{20mm} 
\times 
\Phi^{LL'}_{0} ({\bf k},{\bf k}_+, \omega) = 0
\label{eq7}
\end{eqnarray}
and
\begin{eqnarray}
&& \hspace{-5mm}
\sum_{LL'} \frac{1}{V} \sum_{{\bf k} \sigma} 
\big[\omega J_0^{LL'} ({\bf k},{\bf k}_+) - 
\sum_\beta q_\beta J_\beta^{LL'} ({\bf k},{\bf k}_+)\big] 
\nonumber \\
&& \hspace{15mm} 
\times  
\Phi^{LL'}_{\alpha} ({\bf k},{\bf k}_+, \omega) = 
\sum_\beta q_{\beta} \frac{e^2 n^{\rm tot}_{\beta \alpha}({\bf q})}{m},
\label{eq8}
\end{eqnarray}
where the electron-hole propagator $\Phi^{LL'}_{\nu} ({\bf k},{\bf k}_+, \omega)$ is analytically continued form of 
\begin{eqnarray}
&& \hspace{-5mm}
\Phi^{LL'}_{\nu} ({\bf k},{\bf k}_+,{\it i} \nu_n) = \frac{1}{\beta} \sum_{ {\it i} \omega_m} 
\Phi^{LL'}_{\nu} ({\bf k},{\bf k}_+, {\it i} \omega_n, {\it i} \omega_{n+}).
\label{eq9}
\end{eqnarray}
	In the usual notation for the elements of the RPA irreducible $4 \times 4$ response tensor 
(see Fig.~2)
\begin{eqnarray}
&& \hspace{-10mm}
\pi_{\mu \nu} ({\bf q}, \omega) 
= \sum_{LL'} \frac{1}{V} \sum_{{\bf k} \sigma} J^{LL'}_\mu({\bf k},{\bf k}_+) 
\Phi_{\nu}^{LL'} ({\bf k},{\bf k}_+,\omega),
\label{eq10}
\end{eqnarray}
these two relations can be written as \cite{Kubo95,Kupcic14}
\begin{eqnarray}
&& \hspace{-10mm}
\omega \pi_{0 0}({\bf q}, \omega) = \sum_\alpha q_\alpha \pi_{\alpha 0}({\bf q}, \omega), 
\label{eq11} \\
&& \hspace{-10mm}
\omega \pi_{0 \alpha}({\bf q}, \omega) = \sum_{\beta} q_{\beta}\bigg(\pi_{\beta \alpha}({\bf q}, \omega)  
+ \frac{e^2 n^{\rm tot}_{\beta \alpha}({\bf q})}{m}\bigg).
\label{eq12}
\end{eqnarray}
	This means that Eqs.~(\ref{eq7}) and (\ref{eq8}) 
{{{\rm represent an alternative way to write}}}
the Ward identity relations connecting the renormalized vertices 
$\Gamma^{L'L}_{0}({\bf k}_+,{\bf k},{\it i}\omega_{n+},{\it i}\omega_n)$ and
$\Gamma^{L'L}_{\alpha}({\bf k}_+,{\bf k},{\it i}\omega_{n+},{\it i}\omega_n)$ and thus represent the simplest way to take care of both 
{{{\rm local charge conservation and gauge invariance of the response theory.}}}

\begin{figure}
   \centerline{\includegraphics[width=20pc]{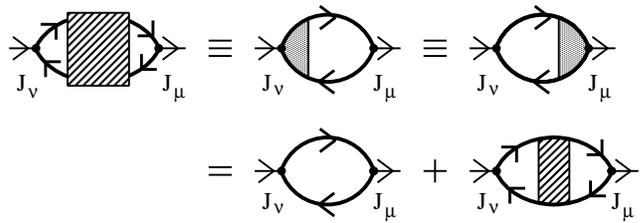}}
   \caption{The Bethe--Salpeter expression for the $ 4 \times 4$ current-current correlation function
   $\pi_{\mu \nu} ({\bf q}, {\it i}\nu_n)$ \cite{Schrieffer64,Kupcic14}. 
   }
  \end{figure}

It is well known that the intraband version of Eqs.~(\ref{eq7}) and (\ref{eq8})  follows directly from 
Eq.~(\ref{eq2}) after multiplication by $J_0^{LL} ({\bf k},{\bf k}_+)$ and summation  over ${\bf k}$ and ${\it i} \omega_n$ \cite{Kupcic15}.
	This means that the intraband Hartree-Fock quantum transport equation is gauge invariant.
	On the other hand, the present form of the interband Hartree-Fock quantum transport equation is not gauge invariant
{{{\rm in the strict sense.}}}
	Namely, after performing the same procedure as in the intraband channel, we obtain extra contributions 
in Eqs.~(\ref{eq7}) and (\ref{eq8}), with an obvious violation of local charge conservation. 
	All these elements become more complicated when there are local field effects \cite{Kupcic13}.
	The restriction to the two-band model for conduction $\pi$ electrons in graphene represents the way 
to obtain a general description of electrodynamic properties of graphene which is still very simple.

\subsection{Dynamical conductivity tensor}
The dynamical conductivity tensor is usually defined by two Kubo formulas  \cite{Kubo95,Kupcic14}
\begin{equation}
\sigma_{\alpha \alpha} ({\bf q},\omega) = \pi_{\alpha \tilde \alpha}({\bf q},\omega) 
= \frac{\it i}{\omega} \bigg(\pi_{\alpha \alpha}({\bf q},\omega)  
+ \frac{e^2 n^{\rm tot}_{\alpha \alpha}({\bf q})}{m}\bigg).
\label{eq13}
\end{equation}
	Here, $\pi_{\alpha \tilde \alpha}({\bf q},\omega)$ is the current-dipole correlation function.
	In the longitudinal case with ${\bf q} = q_\alpha \hat e_\alpha$, it is the product of the current-charge correlation function
$\pi_{\alpha 0}({\bf q},\omega)$ and the dimensionless dipole vertex ${\it i}/q_\alpha$.
	The dipole vertices $ P_\alpha^{LL'} ({\bf k},{\bf k}_+)$ in $\pi_{\alpha \tilde \alpha}({\bf q},\omega)$
are related to the charge vertices from Eq.~(\ref{eq6}) in the following way \cite{Kupcic14}
\begin{eqnarray}
&& \hspace{-10mm}
\sum_\alpha q_\alpha P_\alpha^{LL'} ({\bf k},{\bf k}_+) = {\it i} J_0^{LL'}({\bf k},{\bf k}_+). 
\label{eq14}
\end{eqnarray}
	In the longitudinal case, we can also write
\begin{equation}
\sigma_{\alpha \alpha} ({\bf q},\omega) = \frac{\it i \omega}{q_\alpha^2} \pi_{0 0}({\bf q},\omega).  
\label{eq15}
\end{equation}
	Depending on the complexity of the problem, we can use one of these three expressions for 
$\sigma_{\alpha \alpha} ({\bf q},\omega)$.
	As long as the three types of the correlation function $\pi_{\mu \nu}({\bf q},\omega)$ 
are treated exactly, these three conductivity formulas give the same result.
	Any approximate treatment of the problem usually means that one of these formulas 
is a better choice than the other two.
	In the standard Fermi liquid regime, the current-dipole conductivity formula is the most natural choice.

\subsection{Effective numbers of charge carriers in the partial transverse conductivity sum rule}
The quantity 
\begin{eqnarray}
&& \hspace{-5mm} n^{\rm tot}_{\beta \alpha}({\bf q}) = \sum_{LL'}
\frac{1}{V} \sum_{{\bf k} \sigma} \frac{m}{e^2}
\frac{J_\beta^{LL'} ({\bf k},{\bf k}_+) 
J_{\alpha}^{L'L} ({\bf k}_+,{\bf k})}{\varepsilon_{L'L} ({\bf k}_+,{\bf k})} 
\nonumber \\
&& \hspace{10mm} 
\times [n_L({\bf k}) - n_{L'}({\bf k}_+) ]
\nonumber \\
&& \hspace{7mm} 
= n_{\beta \alpha}^{\rm intra} ({\bf q}) + n_{\beta \alpha}^{\rm inter} ({\bf q})
\label{eq16}
\end{eqnarray}
in Eqs.~(\ref{eq8}), (\ref{eq12}), and (\ref{eq13}) is the total number of charge carriers, which comprises the intraband contribution 
$n_{\beta \alpha}^{\rm intra} ({\bf q})$ ($L=L'$) and the interband contribution $n_{\beta \alpha}^{\rm inter} ({\bf q})$ ($L \neq L'$).
	Since the integrated conductivity spectral weight is proportional to
$n^{\rm tot}_{\beta \alpha}({\bf q}) = (-m/e^2) \pi_{\beta \alpha}({\bf q})$ \cite{Kubo95,Kupcic14}, 
these two numbers can be estimated from the measured intraband and interband 
contributions to the dynamical conductivity $\sigma_{\alpha \alpha} ({\bf q},\omega)$ and compared to the corresponding effective numbers in the
intraband and interband plasmon frequencies (see Sec.~VI). 
	As we show in Sec.~IV,
	{{{\rm there is also a close relation between}}}
$n^{\rm intra}_{\beta \alpha}({\bf q})$, $n^{\rm inter}_{\beta \alpha}({\bf q})$ and two contributions to the effective number 
of charge carriers $n^{\rm dc,tot}_{\beta \alpha}$ in the dc conductivity.
	 $n^{\rm intra}_{\beta \alpha}({\bf q})$ and $n^{\rm inter}_{\beta \alpha}({\bf q})$ are thus 
an important part of the discussion of the mobility of conduction electrons
in doped graphene, as well as in similar multiband electronic systems (Sec.~IV\,B).
	The expression (\ref{eq16}) for the total number of charge carriers is the first important result of the present paper.

\begin{figure}
  \centerline{\includegraphics[width=18pc]{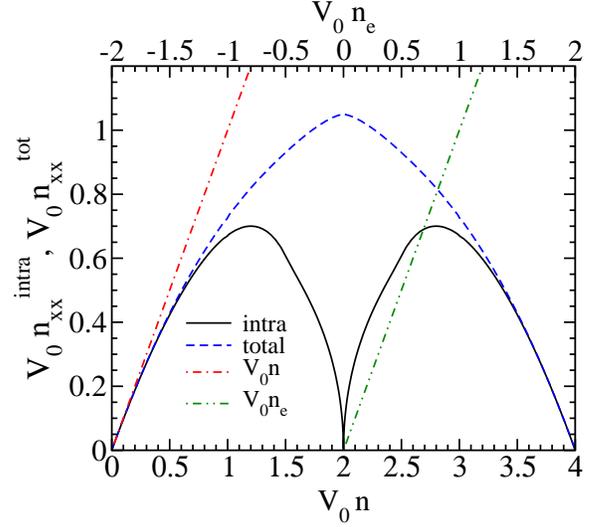}}
   \caption{The effective numbers of charge carriers $n^{\rm intra}_{\alpha \alpha}({\bf q})$ 
   and $n^{\rm tot}_{\alpha \alpha}({\bf q})$ 
   calculated by using Eq.~(\ref{eq16}), for $n_L({\bf k}) = f_L({\bf k})$, $t = 2.52$ eV, $T = 100$ K, and ${\bf q} \approx {\bf 0}$.
   $V_0 n =(1/N) \sum_{L{\bf k} \sigma} f_L ({\bf k})$ is the concentration of conduction electrons
   ($V_0 n=2$ in pristine graphene) and $V_0$ is the primitive cell volume.
}
  \end{figure}  

In the two-band model for $\pi$ electrons in graphene, we have $L = \pi, \pi^*$.
	Figure 3 shows the effective numbers $n^{\rm intra}_{\alpha \alpha}({\bf q})$ and $n^{\rm tot}_{\alpha \alpha}({\bf q})$
obtained by Eq.~(\ref{eq16}), for $n_L({\bf k}) = f_L({\bf k}) \equiv f(\varepsilon_L({\bf k}))$ and ${\bf q} \approx {\bf 0}$.
	The dispersions of electrons in two $\pi$ bands are  \cite{Castro09}
$$
\varepsilon_{\pi^*, \pi} ({\bf k}) = \pm t \sqrt{ 3+ 2 \cos k_xa + 4 \cos \frac{k_x a}{2} \cos \frac{\sqrt{3} k_ya}{2}}.
$$
	Here, $t$ is the first neighbor hopping integral.
	Notice that $n^{\rm tot}_{\alpha \alpha}({\bf q}) \approx n^{\rm intra}_{\alpha \alpha}({\bf q}) \approx n$ for 
$V_0 n < 0.5$, as expected in the usual Fermi liquid regime, and at variance with 
$n^{\rm intra}_{\alpha \alpha}({\bf q}) \propto \sqrt{|n_e|}$ for $V_0 n \approx 2$ 
(here $n_e$ is the concentration of doped electrons/holes measured with respect to the completely occupied $\pi$ band).

{{{\rm It is apparent that}}}
the effective numbers $n^{i}_{\beta \alpha}({\bf q})$, $i = {\rm tot, intra, inter}$, depend on details in the single-electron spectral functions
${\cal A}_L ({\bf k}, \varepsilon)$, but they are not functions of the corresponding electron-hole damping energies.
	This can be easily seen if we show the momentum distribution functions from Eq.~(\ref{eq16}) in their usual explicit form
\begin{eqnarray}
&& \hspace{-10mm}
n_L({\bf k}) = \frac{1}{\beta \hbar} \sum_{{\it i}\omega_n} {\cal G}_L ({\bf k}, {\it i} \omega_n)
\equiv \int_{-\infty}^\infty \frac{d \varepsilon}{2 \pi} \, {\cal A}_L ({\bf k}, \varepsilon) f(\varepsilon).
\label{eq17}
\end{eqnarray}
	Here, $f(\varepsilon)$ is the Fermi-Dirac distribution function and
\begin{eqnarray}
&& \hspace{-5mm}
{\cal A}_L ({\bf k}, \varepsilon) = \frac{\it i}{\hbar} \sum_{s = \pm 1}
s {\cal G}_L ({\bf k}, \varepsilon+ s {\it i} \eta)
\nonumber \\
&& \hspace{10mm} 
= -\frac{2}{\hbar} {\rm Im} \{ {\cal G}_L ({\bf k}, \varepsilon+  {\it i} \eta) \}
\label{eq18}
\end{eqnarray}
is the single-electron spectral function in question.
	Finally,
\begin{eqnarray}
&& \hspace{-10mm} 
{\cal G}_L ({\bf k}, \varepsilon + {\it i} \eta) = \frac{\hbar}{\varepsilon - \varepsilon_L^0({\bf k}) 
- \hbar \Sigma_L ({\bf k},\varepsilon)}
\label{eq19}
\end{eqnarray}
is the $T=0$ retarded single-electron Green's function,
and $\Sigma_L ({\bf k},\varepsilon) \equiv \Sigma_L ({\bf k},\varepsilon + {\it i} \eta)$ is analytically continued 
form of the self-energy (\ref{eq4}).
	The relation (\ref{eq17}) shows a way to incorporate the results of ARPES measurements of ${\cal A}_L ({\bf k}, \varepsilon)$
into the analysis of the dc and dynamical conductivity measurements.

\begin{figure}
  \centerline{\includegraphics[width=17pc]{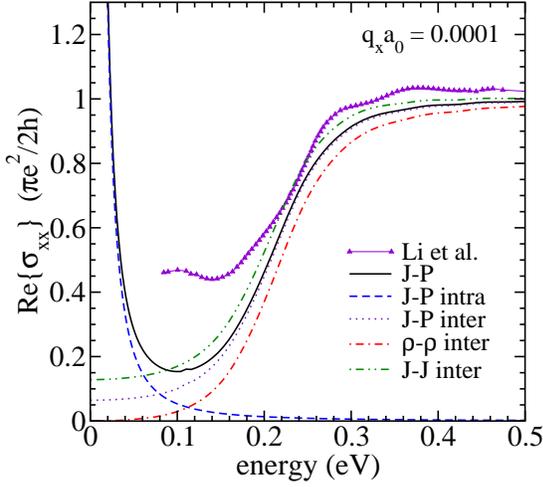}}
   \caption{Solid, dashed, and dotted lines: the real part of the dynamical conductivity of doped graphene obtained 
   by Eq.~(\ref{eq25}) beyond the Dirac cone approximation, 
   for $n_L({\bf k}) = f_{L} ({\bf k})$, $E_{\rm F} = -0.105$ eV, $T=150$ K, $q_xa_0 = 0.0001$, and for realistic values of the damping energies,
   $\hbar \Gamma_1 = 4$ meV and $\hbar \Gamma_2 = 20$ meV.
   The interband part calculated by using $\pi_{00}^{\rm inter} ({\bf q}, \omega)$ and  $\pi_{\alpha0}^{\rm inter} ({\bf q}, \omega)$ from
   Eq.~(\ref{eq24}) is also shown (dot-dashed and dot-dot-dashed lines).
   Experimental data (full triangles) are from Ref.~\cite{Li08}.
   $a_0$ is the Bohr radius
}
  \end{figure}

\subsection{A posteriori relaxation-time approximation} 
In weakly interacting systems, the usual quasi-particle picture
{{{\rm can be safely used,}}}
in which $\varepsilon_{L}^0 ({\bf k}) + \hbar \Sigma_L ({\bf k},\varepsilon)$ in Eq.~(\ref{eq19})
is replaced by $\varepsilon_{L} ({\bf k}) + {\it i} \hbar \Sigma_L^i ({\bf k})$.
	Here, $\Sigma_L^i ({\bf k}) = \Sigma_L^i ({\bf k},\varepsilon=\varepsilon_{L} ({\bf k}))$ is
the single-electron damping energy.
	The result is the spectral function 
\begin{eqnarray}
&& \hspace{-5mm}
{\cal A}_L ({\bf k}, \varepsilon) \approx  \frac{- 2 \hbar \Sigma^i_L ({\bf k})}{
[\varepsilon - \varepsilon_L ({\bf k}) ]^2 + [\hbar \Sigma^i_L ({\bf k})]^2}.
\label{eq20}
\end{eqnarray}
	{{{\rm The next level of approximation}}} 
corresponds to the replacement $n_L({\bf k}) \approx f_L({\bf k})$, i.e., 
$-\Sigma^i_L ({\bf k}) \approx \eta$ in Eq.~(\ref{eq20}).
	This approximation will be referred to as {\it a posteriori} relaxation-time approximation.
	In this case, the total number of charge carriers $n^{\rm tot}_{\beta \alpha}({\bf q})$
is free of any kind of damping effects.

Typical results for the real part of the dynamical conductivity tensor, obtained by using the current-dipole 
approach from Sec.~IV, are shown in Fig.~4 and compared with experimental data.
	The figure illustrates that 
{{{\rm in order to obtain reasonable agreement with experiment}}} 
in the relaxation-time approximation, at least the damping energies $\hbar \Gamma_1$ and $\hbar \Gamma_2$ 
must be treated as independent parameters.
	In the microscopic picture, the difference between $\hbar \Gamma_1$ and $\hbar \Gamma_2$ extracted from measured reflectivity spectra 
reflects the different role of vertex corrections in the intraband and interband quantum transport equations
(notably those related to the long-range Coulomb forces).

The effective numbers associated with $\sigma_{\alpha \alpha} ({\bf q},\omega)$ from Fig.~4 are 
$V_0 n_{\alpha \alpha}^{\rm intra} \approx 23 \times 10^{-3}$ and $V_0 n_{\alpha \alpha}^{\rm inter} \approx 1.027$
(the relation between $n_{\alpha \alpha}^{\rm intra}$ and the nominal concentration of conduction electrons $n$ will be discussed later).
	This means that the effective number $V_0 n_{\alpha \alpha}^{\rm tot} \approx 1.05$ from 
the two-band version of the transverse conductivity sum rule (\ref{eq16}) takes only one half of $V_0n \approx 2$
from the complete transverse conductivity sum rule, in agreement with Fig.~3.

\section{Low-order perturbation theory}
{{{\rm In order to better understand}}}
the microscopic structure of the intraband and interband electron-hole-pair self-energies
[and their imaginary parts ${\it i} \hbar \Gamma_1({\bf k})$ and ${\it i} \hbar \Gamma_2({\bf k})$],
{{{\rm it is helpful to}}}
determine the structure of the $\lambda^0$ and $\lambda^2$ contributions to $\pi_{\mu \nu} ({\bf q}, {\it i}  \nu_n)$,
$\pi^{[0]}_{\mu \nu} ({\bf q}, {\it i}  \nu_n)$ and $\pi^{[2]}_{\mu \nu} ({\bf q}, {\it i}  \nu_n)$
($\lambda$ is the perturbation parameter in $H' = \lambda H_1' + \lambda^2 H_2'$).
	Let us first consider the intraband contributions to $\pi_{\mu \nu} ({\bf q}, {\it i}  \nu_n)$
for electron scattering by phonons.

The calculation of the $\lambda^{0}$ contributions to  $\pi_{\mu \nu}^{\rm intra} ({\bf q}, {\it i}  \nu_n)$
is straightforward. 
	The result contains the factor 
$f_L({\bf k}_+)- f_L({\bf k}) \approx q_\alpha \hbar v_\alpha^L ({\bf k}) \partial f_L({\bf k})/\partial \varepsilon_L^0({\bf k})$
and gives a finite contributions only when multiplied by $1/q_\alpha$ 
[$v_{\alpha}^{L} ({\bf k}) =  J^{LL}_\alpha ({\bf k},{\bf k}) / e = (1/\hbar) \partial \varepsilon_L ({\bf k}) /\partial k_\alpha$
is the electron group velocity].
	Three $\lambda^2$ diagrams give four contributions, which are labeled by $2A_1$, $2A_2$, $2B_1$, and $2B_2$ in Fig.~5.

For example, the result for the  $2A_1$ diagram is given by
\begin{eqnarray}
&& \hspace{-5mm}
\pi^{[2A_1]}_{\mu \nu} ({\bf q}, {\it i}  \nu_n) 
= \sum_{LL'} \frac{1}{V} \sum_{{\bf k} \sigma} J_\mu^{LL'}({\bf k},{\bf k}_+) J_\nu^{L'L}({\bf k}_+,{\bf k})
\nonumber \\
&&  \hspace{5mm}
\times  \sum_{\lambda {\bf k'}} \frac{ |G_\lambda({\bf k},{\bf k}')|^2}{N}
\sum_{s = \pm 1} s {\cal S}^{[2A_1]}({\it i}  \nu_n, \varepsilon, \varepsilon_+,\varepsilon',s\omega'), 
\nonumber \\
\label{eq21}
\end{eqnarray}
with $L'=L$.
	Here, ${\cal S}^{[2A_1]}({\it i}  \nu_n, \varepsilon, \varepsilon_+,\varepsilon',\omega')$ is the corresponding Matsubara sum,
which has the following structure
\begin{eqnarray}
&& \hspace{-10mm}
{\cal S}^{[2A_1]}({\it i}  \nu_n, \varepsilon, \varepsilon_+,\varepsilon',\omega')
\nonumber \\
&&  \hspace{5mm}
=\frac{f^b (\omega') + f(\varepsilon')}{{\it i} \hbar \nu_n  + \varepsilon- \varepsilon' + \hbar \omega'} 
\frac{f(\varepsilon)-f(\varepsilon_+)}{({\it i} \hbar \nu_n  + \varepsilon- \varepsilon_+)^2}
\nonumber \\
&&  \hspace{8mm}
+\frac{f^b (\omega') + f(\varepsilon')}{{\it i} \hbar \nu_n  + \varepsilon- \varepsilon' + \hbar \omega'} 
\frac{f(\varepsilon)-f(\varepsilon'-\hbar \omega')}{(\varepsilon_+- \varepsilon' + \hbar \omega')^2}
\nonumber \\
&&  \hspace{8mm}
-\frac{f^b (\omega') + f(\varepsilon')}{\varepsilon_+- \varepsilon' + \hbar \omega'} 
\frac{\partial f(\varepsilon_+)/\partial \varepsilon_+}{{\it i} \hbar \nu_n  + \varepsilon- \varepsilon_+}
\label{eq22}
\end{eqnarray}
(similarly for the $2A_2$, $2B_1$, and $2B_2$ sums).
	Here, 
$\varepsilon = \varepsilon_L({\bf k})$,  $\varepsilon_+ = \varepsilon_{L'}({\bf k}_+)$, 
$\varepsilon' = \varepsilon_{L'}({\bf k}_+')$, $\omega' = \omega_{\lambda {\bf k}'-{\bf k}}$,
$G_\lambda({\bf k},{\bf k}')$ is the electron-phonon coupling function,
$\omega_{\lambda {\bf q}}$ is the phonon frequency, and $f^b (\omega')$ is the Bose-Einstein distribution function.

\begin{figure}[tb]
   \centerline{\includegraphics[width=18pc]{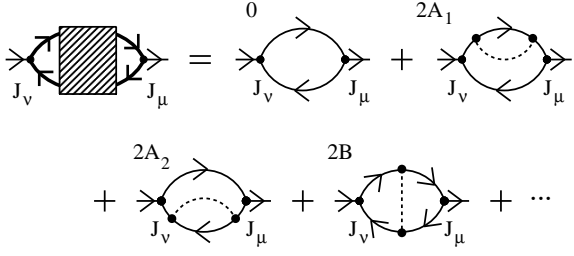}}
   \caption{One $(H_1')^0$ and three $(H_1')^2$ contributions to $\pi_{\mu \nu} ({\bf q}, {\it i}  \nu_n) $, labeled 
   by 0 (bare contribution), $2A_1$ (electron self-energy term), $2A_2$ (hole self-energy term), and $2B=2B_1+2B_2$ (vertex correction).
   }
   \end{figure}

The four Matsubara sums in $\pi^{\rm intra [2]}_{\mu \nu} ({\bf q}, {\it i}  \nu_n)$ comprise three different contributions
associated with three terms in Eq.~(\ref{eq22}).
	The direct contributions are characterized by the factor $f(\varepsilon_+)- f(\varepsilon) \propto q_\alpha$,
while the indirect contributions are proportional to $f(\varepsilon)-f(\varepsilon'-\hbar \omega')$.
	The latter give rise to a product of two effective vertex functions of the form
$[J_\mu^{LL}({\bf k},{\bf k}_+)- J_\mu^{LL}({\bf k}',{\bf k}'_+)][J_\nu^{LL}({\bf k}_+,{\bf k})-J_\nu^{LL}({\bf k}'_+,{\bf k})']$, 
which vanishes  for $\mu=0$ and/or $\nu =0$, because $J_0^{LL}({\bf k},{\bf k}_+) \approx e$ \cite{Kupcic07}.
	The third term is related to the renormalization of the electron dispersions in $f_L({\bf k})$ and $f_{L'}({\bf k}_+)$,
and does not appear in the vertex corrections contribution.
	The recollection of the diagrams of the third type in powers of $\lambda^2$ to infinity leads in a natural way 
to the momentum distribution function $n_L({\bf k})$ from Eq.~(\ref{eq17}).
	
There is a well-defined exclusion rule, which 
{{{\rm is a direct consequence of}}} 
the identity relation (\ref{eq11}).
	The direct contributions are relevant only to the correlation functions $\pi^{\rm intra [2]}_{\mu \nu} ({\bf q}, {\it i}  \nu_n)$ 
in which at least one vertex is the charge vertex,
leading, for example, to the usual expressions for the ${\bf k}$-dependent intraband memory function \cite{Kupcic15}.
	Their contribution to the current-current correlation function
$\pi_{\alpha \alpha}^{\rm intra} ({\bf q}, \omega)$ is thus negligible, due to the factor $q_\alpha^2$.
	In $\pi_{\alpha \alpha}^{\rm intra} ({\bf q}, \omega)$, the leading role is played by the indirect contributions.

There is no such rule for the interband contributions.
	The $\lambda^0$ contribution $\pi_{\alpha \alpha}^{\rm inter [0]} ({\bf q}, \omega)$, given by the $L \neq L'$
contributions in 
\begin{eqnarray}
&& \hspace{-7mm} 
\pi_{\alpha \alpha}^{[0]} ({\bf q}, \omega) =
\sum_{L L'} \frac{1}{V} \sum_{{\bf k} \sigma} |J^{LL'}_{\alpha}({\bf k},{\bf k}_+)|^2
\frac{f_{L'}({\bf k}_+)-f_L({\bf k})}{\hbar \omega + \varepsilon_{LL'}({\bf k},{\bf k}_+) + {\it i}\eta},
\nonumber \\
\label{eq23}
\end{eqnarray}
is finite and it is directly related to $\pi_{\alpha 0}^{\rm inter [0]} ({\bf q}, \omega)$.
	The relaxation processes in $\pi_{\mu \nu}^{\rm inter} ({\bf q}, \omega)$, which start with the $\lambda^2$ contributions, lead thus 
to the redistribution of the spectral weight over a slightly wider energy range than in $\pi_{\mu \nu}^{\rm inter [0]} ({\bf q}, \omega)$.
	In the leading approximation, we obtain
\begin{eqnarray}
&& \hspace{-0mm} 
\pi_{\mu \nu}^{\rm inter} ({\bf q}, \omega) \approx
\sum_{L \neq L'} \frac{1}{V} \sum_{{\bf k} \sigma} J^{LL'}_{\mu}({\bf k},{\bf k}_+) 
\Phi_{\nu}^{LL'} ({\bf k},{\bf k}_+,\omega)
\nonumber \\
&& \hspace{-5mm} 
\Phi_{\nu}^{LL'} ({\bf k},{\bf k}_+,\omega) = 
\frac{J^{L'L}_{\nu}({\bf k}_+,{\bf k})[n_{L'}({\bf k}_+) - n_L({\bf k})]}{
\hbar \omega + \varepsilon_{LL'}({\bf k},{\bf k}_+) + {\it i}\hbar \Gamma^{LL'} ({\bf k})},
\label{eq24}
\end{eqnarray}
where the $\Gamma^{LL'} ({\bf k})$ are the damping energies in question
[$\Gamma^{LL} ({\bf k}) \approx \Gamma_1$ and $\Gamma^{L\underline L} ({\bf k}) \approx \Gamma_2$ in Fig.~4].

{{{\rm The dot-dashed and the dot-dot-dashed lines}}} 
in Fig.~4 show the predictions of the current-current 
and the charge-charge conductivity formulas for the interband dynamical conductivity [given, respectively, 
by the second expression in Eq.~(\ref{eq13}) and by Eq.~(\ref{eq15})].
	There is quite a large difference between the three interband contributions at $\omega \approx 0$.
	The charge-charge conductivity formula underestimates the interband contribution to the $\omega \approx 0$  conductivity
and the current-current contribution overestimates it.
	Finally, it should be noticed that in spite of the fact that the integrated interband spectral weight is almost the same for the three cases, 
it is obvious that only the current-dipole conductivity formula gives the result which is identical to 
the general expression for the partial transverse conductivity sum rule from Eq.~(\ref{eq16}).

\section{Longitudinal current-dipole approach}
For many purposes it is sufficient to use the semiclassical version of Eq.~(\ref{eq2}) in which the relaxation processes
associated with the interactions in $H_1'$ and $H_2'$
are described in terms of the intraband and interband memory functions $M^{LL'}_\alpha ({\bf k},\omega)$.
	The result is the current-dipole conductivity formula \cite{Kupcic14,Kupcic15}
\begin{eqnarray}
&& \hspace{-10mm} 
\sigma_{\alpha \alpha} ({\bf q}, \omega) =
\sum_{LL'} \frac{1}{V} \sum_{{\bf k} \sigma}
\frac{{\it i} \hbar |J^{LL'}_{\alpha}({\bf k},{\bf k}_+)|^2}{\varepsilon_{LL'}({\bf k},{\bf k}_+)}
\nonumber \\
&& \hspace{5mm}
\times \frac{n_{L'}({\bf k}_+)-n_L({\bf k})}{\hbar \omega + \varepsilon_{LL'}({\bf k},{\bf k}_+) + {\it i}\hbar \Gamma^{LL'}({\bf k})},
\label{eq25}
\end{eqnarray}
which consists of the interband contribution $\pi_{\alpha 0}^{\rm inter} ({\bf q}, \omega)$ from Eq.~(\ref{eq24}) and
the analogous expression for $\pi_{\alpha 0}^{\rm intra} ({\bf q}, \omega)$.
	The $\hbar \Gamma^{LL'}({\bf k})$ are the intraband and interband electron-hole-pair damping energies, which 
are proportional to the imaginary part of the memory functions 
$M^{LL'}_\alpha ({\bf k}, \omega)$ taken at $\hbar \omega =\varepsilon_{L'L}({\bf k}_+,{\bf k})$.
	The exclusion rule from the previous section is implicitly included through the very definition 
of the intraband memory functions $M^{LL}_\alpha ({\bf k}, \omega)$.
	However, to estimate $M^{LL'}_\alpha ({\bf k}, \omega)$, $L \neq L'$, we must solve the self-consistent 
integral equation (\ref{eq2}) [or Eq.~(\ref{eq1})].
	The limit $\Gamma^{LL} ({\bf k}) \approx \Gamma_1$, $\Gamma^{L\underline L} ({\bf k}) \approx \Gamma_2$
corresponds to the aforementioned {\it a posteriori} relaxation-time approximation, with $n_L({\bf k}) \neq f_L({\bf k})$.

\subsection{DC conductivity of lightly doped graphene}
{{{\rm At the level of approximation used in Eq.~(\ref{eq25}),}}} 
the dc conductivity of the two-band model for $\pi$ electrons in graphene becomes
\begin{eqnarray}
&& \hspace{-5mm} 
\sigma_{\alpha \alpha}^{\rm dc} =
\sum_{LL'} \frac{1}{V} \sum_{{\bf k} \sigma}
\frac{\hbar |J^{LL'}_{\alpha}({\bf k},{\bf k}_+)|^2}{\varepsilon_{LL'}({\bf k},{\bf k}_+)}[n_{L'}({\bf k}_+)-n_L({\bf k})]
\nonumber \\
&& \hspace{5mm}
\times \frac{\hbar \Gamma^{LL'}({\bf k})}{\varepsilon_{LL'}^2({\bf k},{\bf k}_+) + [\hbar \Gamma^{LL'}({\bf k})]^2}.
\label{eq26}
\end{eqnarray}
	For $\Gamma^{LL}({\bf k}) = \Gamma_1$ and $\Gamma^{L\underline{L}}({\bf k}) = \Gamma_2$, 	
the intraband contribution to $\sigma_{\alpha \alpha}^{\rm dc}$,
\begin{eqnarray}
&& \hspace{-10mm} 
\sigma_{\alpha \alpha}^{\rm dc, intra} = \frac{e^2}{\Gamma_1} 
\frac{1}{V} \sum_{L {\bf k} \sigma} [v^{L}_{\alpha}({\bf k})]^2
\frac{n_{L}({\bf k}_+)-n_L({\bf k})}{\varepsilon_{LL}({\bf k},{\bf k}_+)},
\label{eq27}
\end{eqnarray}
is the product of the relaxation time $\tau_1 = 1/\Gamma_1$ and the intraband part of the total number of charge carriers
\begin{eqnarray}
&& \hspace{-5mm} n^{\rm intra}_{\alpha \alpha} = \frac{1}{V} \sum_{L{\bf k} \sigma} m 
[v_\alpha^{L} ({\bf k})]^2
\bigg( -\frac{\partial n_L({\bf k})}{\partial \varepsilon_L({\bf k})}\bigg).
\label{eq28}
\end{eqnarray}
	The latter has the same structure as $n^{\rm intra}_{\alpha \alpha} ({\bf q}\approx {\bf 0})$  
from the partial transverse conductivity sum rule (\ref{eq16}).

On the other hand, the interband contribution reads 
\begin{eqnarray}
&& \hspace{-10mm} 
\sigma_{\alpha \alpha}^{\rm dc, inter} = \frac{e^2}{m \Gamma_2} n^{\rm dc, inter}_{\alpha \alpha}
\label{eq29} \\
&& \hspace{-10mm} n^{\rm dc, inter}_{\alpha \alpha} = 
\frac{1}{V} \sum_{L {\bf k} \sigma}
\frac{m}{e^2} |J^{L\underline{L}}_{\alpha}({\bf k},{\bf k}_+)|^2
\nonumber \\
&& \hspace{6mm} \times  \frac{n_{\underline{L}}({\bf k}_+)-n_L({\bf k})}{
\varepsilon_{L\underline{L}}({\bf k},{\bf k}_+)}
\frac{(\hbar \Gamma_2)^2}{\varepsilon_{L\underline{L}}^2({\bf k},{\bf k}_+) + (\hbar \Gamma_2)^2}.
\label{eq30}
\end{eqnarray}
	It should be noticed that $n^{\rm dc, inter}_{\alpha \alpha}$ represents a small fraction 
of $n^{\rm inter}_{\alpha \alpha}({\bf q}\approx {\bf 0})$
from Eq.~(\ref{eq16}), which is selected by the function 
$(\hbar \Gamma_2)^2/[\varepsilon_{L\underline{L}}^2({\bf k},{\bf k}_+) + (\hbar \Gamma_2)^2]$.
	This means that only the states in the vicinity of the Fermi level satisfying the condition
$\varepsilon_{L\underline{L}}^2({\bf k},{\bf k}_+) < (\hbar \Gamma_2)^2$ participate in the interband dc conductivity.
	This term is negligible in usual multiband electronic systems, but it is finite in graphene and in similar systems
with negligible threshold energy for interband electron-hole excitations.
	It must also be noticed that although the analysis of $n^{\rm inter}_{\alpha \alpha}({\bf q})$ 
requires the treatment of the interband electron-hole excitations beyond the Dirac cone approximation. this approximation 
{{{\rm can safely be used}}}
in analyzing $n^{\rm dc, inter}_{\alpha \alpha}$ and $n^{\rm intra}_{\alpha \alpha}$.

\begin{figure}
  \centerline{\includegraphics[width=17pc]{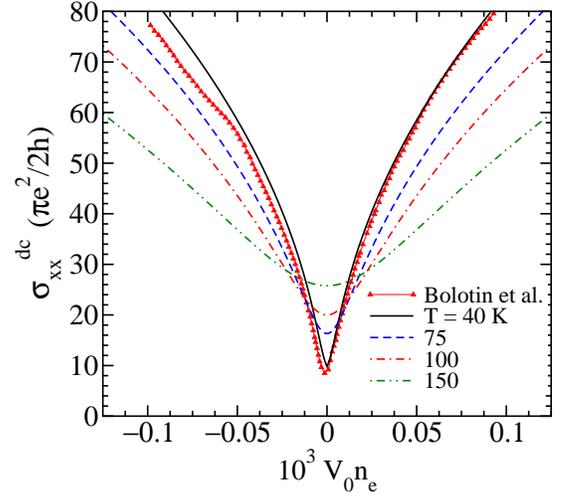}}
   \caption{The doping dependence of $\sigma_{\alpha \alpha}^{\rm dc}$ in ultraclean graphene calculated by using 
   the current-dipole conductivity formula (\ref{eq25}),
   for $n_L({\bf k}) = f_{L} ({\bf k})$, and
   for realistic values of $\hbar \Gamma$, $\hbar \Gamma(T) = a + b T$, $a =0.5$  meV and $b = 0.5/200$ meV/K.
   Experimental data, taken at $T = 40$ K, are from Ref.~\cite{Bolotin08}.
}
  \end{figure}

Figure 6 shows the doping dependence of the dc conductivity in ultraclean graphene at temperatures between 40 K and 150 K.
	The calculation is performed in the Dirac cone approximation, by using the usual replacement 
for the square of the current vertices \cite{Ando98},
\begin{eqnarray}
&& \hspace{-10mm}
|J^{LL'}_{\alpha}({\bf k},{\bf k}_+)|^2 \rightarrow
\frac{1}{2} \sum_\alpha |J^{LL'}_{\alpha}({\bf k},{\bf k}_+)|^2 = \frac{1}{2} (ev_{\rm F})^2.
\label{eq31}
\end{eqnarray}
	The relaxation rates are taken to be $\hbar \Gamma_1 = \hbar \Gamma_2 = \hbar \Gamma (T)= a + b T$, for simplicity,
where $a$ and $b$ are functions of Fermi energy \cite{Ando98,Peres06,Bolotin08}.
	The damping energy $-\hbar \Sigma^i_L ({\bf k})$ in Eq.~(\ref{eq20}) is approximated by $\hbar \Sigma^i = 0$,
leading to $n_L({\bf k}) = f_L({\bf k})$.
	Notice that for $\hbar \Gamma_1 = \hbar \Gamma_2 = \hbar \Gamma$, we can introduce the effective number
of charge carriers  $n^{\rm dc, tot}_{\alpha \alpha}$ in the dc conductivity, which is proportional to $\sigma_{\alpha \alpha}^{\rm dc}$,
\begin{eqnarray}
&& \hspace{-10mm} 
V_0n^{\rm dc, tot}_{\alpha \alpha} = \frac{V_0 m}{\hbar e^2} \hbar \Gamma \sigma_{\alpha \alpha}^{\rm dc}
= \bigg(\frac{2 h}{\pi e^2}\bigg) \frac{\sqrt{3}}{4 t} \hbar \Gamma \sigma_{\alpha \alpha}^{\rm dc}.
\label{eq32}
\end{eqnarray}

\begin{figure}
  \centerline{\includegraphics[width=20pc]{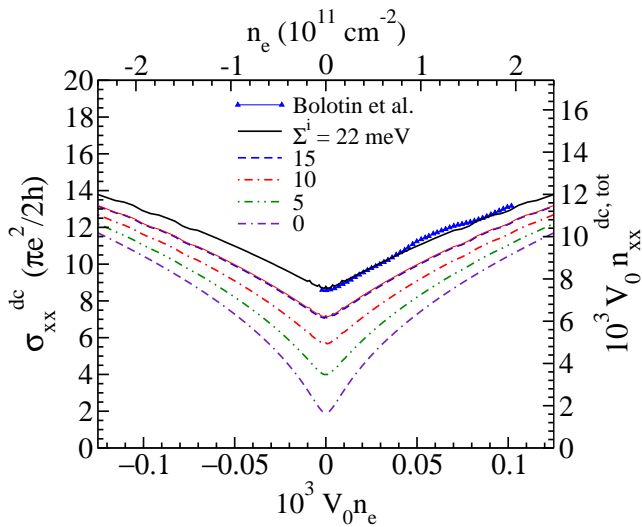}}
   \caption{The doping dependence of the effective number $n^{\rm dc, tot}_{\alpha \alpha}$ and the dc conductivity $\sigma_{\alpha \alpha}^{\rm dc}$
   for different values of the ratio $\Sigma^i/\Gamma$ at $T=50$ K and $\hbar \Gamma = 5$ meV.
   Experimental data (for $n_e > 0$ and $T = 40$ K) are from Ref.~\cite{Bolotin08}.
}
  \end{figure} 

Figure 7 illustrates the doping dependence of $\sigma_{\alpha \alpha}^{\rm dc}$ in dirty graphene at $T = 50$ K for different values of the ratio
$\Sigma^i/\Gamma$.
	{{{\rm The result is in reasonably good agreement with experiment}}} 
for $\hbar \Gamma = 5$ meV and $ \hbar \Sigma^i = 22$ meV.
	Notice that the dependence of $\sigma_{\alpha \alpha}^{\rm dc}$ (and $n^{\rm dc, tot}_{\alpha \alpha}$) on $|n_e|$ changes from the linear dependence for 
$\Sigma^i/\Gamma \gg 1$ to the $\sqrt{|n_e|}$ dependence in the opposite limit.
	Therefore Figs.~6 and 7 show that the main effects of current annealing of the samples \cite{Bolotin08}
are the reduction of the damping energies $\hbar \Gamma_1$ and
$\hbar \Gamma_2$ by one order of magnitude and a much larger effect on the ${\bf q} \approx {\bf 0}$ scattering processes in $\hbar \Sigma^i$.
	Figure 7 also illustrates the dependence of $n^{\rm dc, tot}_{\alpha \alpha}$ 
on the single-electron damping energy $\Sigma^i$ in the ballistic conductivity regime of dirty graphene samples.

In conclusion, to understand the damping effects in dirty conductors quantitatively, 
we have to take into account vertex corrections in the quantum transport equations, at least in the phenomenological way.
	There are significant contributions in the damping energies $\hbar \Sigma^i_L ({\bf k})$ originating 
from the ${\bf q} \approx {\bf 0}$ forward scattering processes, which are canceled out 
in the electron-hole damping energies $\Gamma^{LL'}({\bf k})$, resulting in the regime $2\Sigma^i/\Gamma \gg 1$ shown in Fig.~7.

\subsection{Mobility of conduction electrons}
It is also important to recall that the mobility of conduction electrons $\mu$ is the quantity which is intimately related with
the effective number of charge carriers.
	In simple semiconducting systems, the mobility is usually defined by \cite{Ziman72}
\begin{eqnarray}
&& \hspace{-10mm} 
\sigma_{\alpha \alpha}^{\rm dc} =
\sigma_{\alpha \alpha}^{\rm dc, intra} + \sigma_{\alpha \alpha}^{\rm dc, inter} = e \mu |n_e|,
\label{eq33}
\end{eqnarray}
where $|n_e|$ is the nominal concentration of doped conduction electrons/holes.
	Figure 7 illustrates that in spite of the fact that 
this definition of $\mu$ is widely used in analyzing experimental results in graphene \cite{Novoselov05,Zhang05}
(for example, $\mu$ is estimated to be as large as $170\,000$ cm$^2/$V at $n = 2 \times 10^{11}$ cm$^{-2}$ \cite{Bolotin08})
it makes sense only for large enough Fermi energies (typically $V_0|n_e| > 10^{-2}$).
	For example, when $10^3 V_0 n^{\rm dc, tot}_{\alpha \alpha} \approx 10$ from Fig.~7 is replaced by
$10^3 V_0 n_e \approx 0.1$, the mobility $\mu$ increases by two orders of magnitude with respect to the correct value 
$\mu = (e/m \Gamma)$.
	It increases further with decreasing $|n_e|$ and becomes infinite at $n_e =0$.
	The mobility
{{{\rm that is infinite is certainly not physically reasonable.}}}

A more realistic form of $\sigma_{\alpha \alpha}^{\rm dc}$
treats the intraband and interband contributions in Eq.~(\ref{eq26}) as two independent terms characterized by two mobilities,
$\mu^{\rm intra} = (e/m \Gamma_1)$ and $\mu^{\rm dc, inter} = (e/m \Gamma_2)$.
	In this case, we obtain the general form of the dc conductivity in graphene,
\begin{eqnarray}
&& \hspace{-10mm} 
\sigma_{\alpha \alpha}^{\rm dc} = e \mu^{\rm intra} n^{\rm intra}_{\alpha \alpha} +
e \mu^{\rm dc, inter} n^{\rm dc, inter}_{\alpha \alpha}.
\label{eq34}
\end{eqnarray}
	It is
{{{\rm very much reminiscent of}}} 
the dc conductivity of the two-band semiconductors \cite{Ziman72}.

\section{Transverse current-dipole approach}
{{{\rm It is tempting to}}} 
use the procedure of calculating the current-current conductivity formula (\ref{eqA2}) from Appendix A \cite{Mahan90,Ando98}
{{{\rm to work out}}} 
the other elements of the $4 \times 4$ response tensor.
	In this way it is possible to obtain an alternative form of the current-dipole conductivity formula, 
which can be useful when comparing the results of the present paper 
{{{\rm with previous work,}}}
in particular with that based on the current-current approach \cite{Peres08,Carbotte10}.

The result, 
\begin{eqnarray}
&& \hspace{-2mm} 
\pi_{\mu \nu} ({\bf q}, \omega) = 
\sum_{LL'} \frac{1}{V} \sum_{{\bf k} \sigma}
J^{LL'}_{\mu}({\bf k},{\bf k}_+)  \Phi_{\nu}^{LL'} ({\bf k},{\bf k}_+,\omega),
\nonumber \\
&& \hspace{-10mm}
\Phi_{\nu}^{LL'} ({\bf k},{\bf k}_+,\omega) = 
 \int_{-\infty}^\infty \frac{ d \varepsilon}{2 \pi} \int_{-\infty}^\infty \frac{ d \varepsilon'}{2 \pi}
{\cal A}_{L} ({\bf k}, \varepsilon) {\cal A}_{L} ({\bf k}_+, \varepsilon') 
\nonumber \\
&& \hspace{15mm}
\times J^{L'L}_{\nu}({\bf k}_+,{\bf k}) \frac{f(\varepsilon)-f(\varepsilon')}{\hbar \omega  +{\it i} \eta + \varepsilon - \varepsilon'},
\label{eq35}
\end{eqnarray}
is characterized by the product of two Lorentz functions, ${\cal A}_{L} ({\bf k}, \varepsilon)$ and
${\cal A}_{L'} ({\bf k}_+, \varepsilon')$, and the function $[\hbar \omega + \varepsilon - \varepsilon' +{\it i} \eta]^{-1}$.
	The ideal conductivity regime in ${\rm Im} \{ \pi_{\mu \nu}^{\rm intra} ({\bf q}, \omega) \}$
[${\cal A}_{L} ({\bf k}, \varepsilon) \approx {\cal A}_{L}^0  ({\bf k}, \varepsilon)= 2 \pi \delta(\varepsilon-\varepsilon_L({\bf k}))$,
in this case]
leads to a longstanding problem of the product of three $\delta$-functions.
	The conductivity formula (\ref{eqA2}) in Appendix A is obtained by using the function $\delta (\hbar \omega + \varepsilon - \varepsilon')$
to evaluate the integral over $\varepsilon'$ and then integrating over $\varepsilon$.

\begin{figure}
  \centerline{\includegraphics[width=17pc]{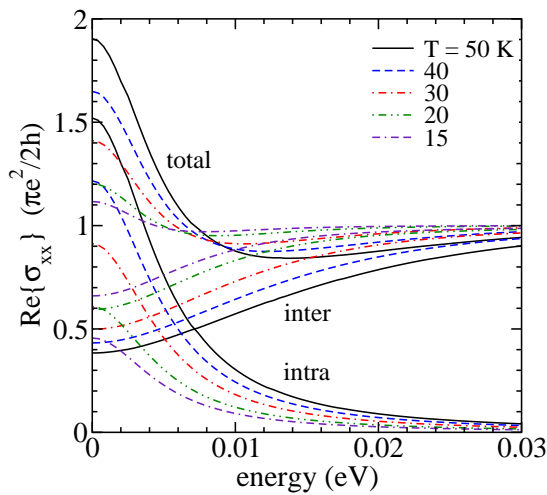}}
   \caption{The real part of the dynamical conductivity of pristine graphene calculated by using Eq.~(\ref{eq25}), 
   for $n_L({\bf k}) = f_{L} ({\bf k})$,  $\hbar \Gamma_1=  \hbar \Gamma_2= 5$ meV, and $T=50$, 40, 30, 20, 15 K.
   The intraband and interband contributions are also shown.
}
  \end{figure}

The same order of steps in evaluating $\pi_{\alpha 0}^{\rm intra} ({\bf q}, \omega)$ leads to the result which is proportional to $\omega / q_\alpha$.
	This result is evidently incorrect, because it is singular in the Drude limit $\omega^2 \gg q_\alpha^2 [v_\alpha^L ({\bf k})]^2$.
	Evidently, to obtain an alternative form of Eq.~(\ref{eq25}),  which is correct in both the intraband and interband channel, the product 
${\cal A}_{L} ({\bf k}, \varepsilon) {\cal A}_{L'} ({\bf k}_+, \varepsilon') [f(\varepsilon)-f(\varepsilon')]$ must be replaced by 
${\cal A}_{L} ({\bf k}, \varepsilon) {\cal A}_{L'} ({\bf k}_+, \varepsilon') [f_L({\bf k})-f_{L'}({\bf k}_+)]$.
	The result is
\begin{eqnarray}
&& \hspace{-10mm} 
{\rm Re} \{ \sigma_{\alpha \alpha}^0 ({\bf q}, \omega) \} =
\int_{-\infty}^\infty \frac{ d \varepsilon}{4 \pi}
\sum_{LL'} \frac{1}{V} \sum_{{\bf k} \sigma} \hbar |J^{LL'}_{\alpha}({\bf k},{\bf k}_+)|^2  
\nonumber \\
&& \hspace{7mm}
\times {\cal A}_{L} ({\bf k}, \varepsilon) {\cal A}_{L'} ({\bf k}_+, \varepsilon_+)
\frac{f_{L} ({\bf k}) - f_{L'} ({\bf k}_+)}{\varepsilon_{L'L}({\bf k}_+,{\bf k})}.
\label{eq36}
\end{eqnarray}
The simplest way to verify this result analytically is to compare the predictions for $\pi_{\alpha 0}^{\rm intra[2]} ({\bf q}, \omega)$
with the results of 
{{{\rm low-order perturbation theory}}}
from Sec.~III.

This expression for the real part of the dynamical conductivity is the second important result of the present paper.
	{{{\rm We can therefore conclude that}}} 
the vertex corrections are not only 
{{{\rm an essential part of}}}
the aforementioned exclusion rule, but also represents a criterion how to deal with the factor $f(\varepsilon)-f(\varepsilon')$ in Eq.~(\ref{eq35})
and in similar expressions.

Figure 8 shows the results for ${\rm Re} \{ \sigma_{\alpha \alpha} (\omega) \}$ in pristine graphene obtained 
by two current-dipole conductivity formulas, 
Eqs.~(\ref{eq25}) and (\ref{eq36}), in the relaxation-time approximation, with $n_L({\bf k}) = f_{L} ({\bf k})$.
	The result is the same for both approaches.
	Notice that at $T=0$ K the intraband contribution vanishes, as well as that the interband one is characterized by the well-known value
$(\pi e^2/ 2 h)$ \cite{Ziegler06}.

The advantages of Eq.~(\ref{eq25}) over Eq.~(\ref{eq36}) [and over the usual current-current conductivity formula (\ref{eqA2})] are obvious.
	This formula treats the damping energies $\hbar \Gamma_1$, $\hbar \Gamma_2$, and
$\hbar \Sigma^i$ as three independent parameters.
	It anticipates the effects of vertex corrections and is thus capable of explaining the relation between $\hbar \Sigma^i$
estimated from measured ARPES spectra \cite{Pletikosic12} and $\hbar \Gamma_1$, $\hbar \Gamma_2$ 
extracted from reflectivity and dc measurements \cite{Bolotin08,Li08}.
	As mentioned above, the $\hbar \Sigma^i \neq 0$ regime in Eq.~(\ref{eq25}) characterizes dirty graphene samples.
	It is even more important in analyzing different strongly correlated systems such as underdoped cuprates \cite{KupcicUP}.
	This is the third major conclusion of the present analysis.

\begin{figure}
  \centerline{\includegraphics[width=17pc]{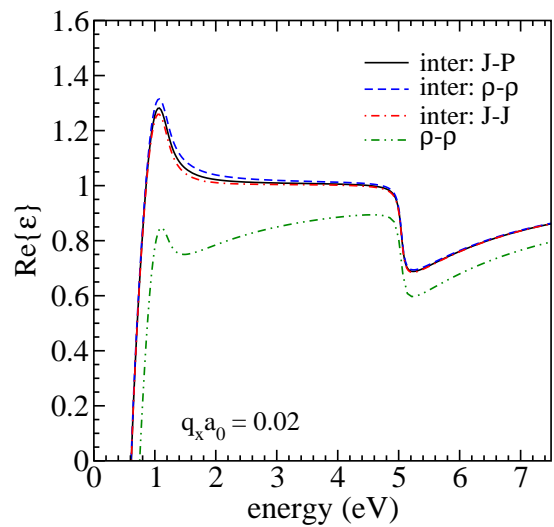}}
   \caption{The real part of $\varepsilon({\bf q}, \omega)$ calculated by using three versions of 
   $\pi^{\rm inter}_{\mu \nu} ({\bf q}, \omega)$ from Eq.~(\ref{eq24}), for $E_{\rm F} = -0.5$ eV and $q_x a_0 = 0.02$, $q_y=0$.
   $\sigma_{\alpha \alpha}^{\rm intra} ({\bf q},\omega)$ is given by the intraband term in Eq.~(\ref{eq25}).
   The dot-dot-dashed line is the prediction of the charge-charge conductivity formula.
   The parameters of the model are $\hbar \Gamma_1 = 10$ meV, $\hbar \Gamma_2 = 50$ meV, and $T=150$ K.
}
  \end{figure}

\section{Dirac and $\pi$ plasmons}
In Sec.~III, we have seen  that the problem with phenomenological treatment of the relaxation processes 
in the intraband charge-charge correlation function can be solved by recollecting the diagrams associated 
with the scattering processes in $H' = \lambda H_1' + \lambda^2 H_2'$ in powers of $\lambda^2$ to infinity.
	However, 
{{{\rm to do this, we}}} 
must take care of Eq.~(\ref{eq11}); otherwise, the local charge will not be conserved. 
	The violation of local charge conservation is expected to be visible in both 
the low-frequency conductivity (as already shown in Fig.~4) and in the dispersion of the intraband plasmon resonance.

Figure 9 illustrates the real part of the dielectric function for $E_{\rm F} = -0.5$ eV and $q_x a_0 = 0.02$ 
obtained by combining the current-dipole expression for 
$\sigma_{\alpha \alpha}^{\rm intra} ({\bf q},\omega)$ from Eq.~(\ref{eq25}) with three expressions for 
$\sigma_{\alpha \alpha}^{\rm inter} ({\bf q},\omega)$ from Eq.~(\ref{eq24}).
	The Dirac plasmon frequency $\omega_{\rm pl} ({\bf q})$ is essentially the same for all three cases.
	On the other hand, the charge-charge version of 
$\sigma_{\alpha \alpha}^{\rm tot} ({\bf q},\omega) = 
\sigma_{\alpha \alpha}^{\rm intra} ({\bf q},\omega)+\sigma_{\alpha \alpha}^{\rm inter} ({\bf q},\omega)$,
which is widely used in analyzing interband collective modes \cite{Despoja13,Novko15,Liou15},
leads to a small shift of $\omega_{\rm pl} ({\bf q})$ to higher frequencies.

 \begin{figure}
  \centerline{\includegraphics[width=17pc]{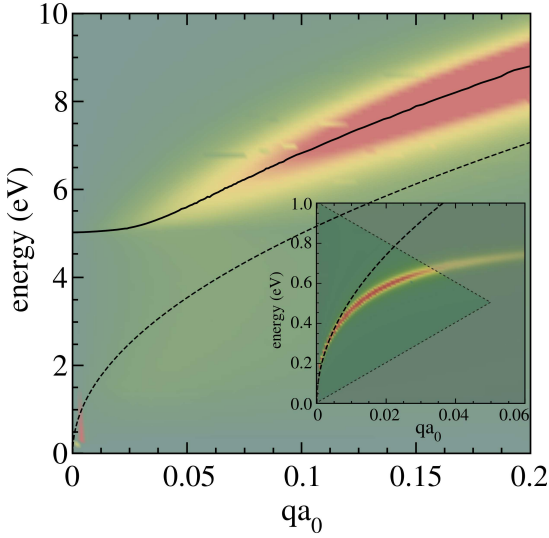}}
   \caption{The two-dimensional plot of the energy loss function $-{\rm Im} \{ 1/\varepsilon ({\bf q}, \omega) \}$,
   for $q_x=q$ and $q_y = 0$.
   Main figure: the interband plasmon resonance ($\pi$ plasmon, 
   {{{\rm in common language).}}}
   Inset: the intraband (Dirac) plasmon resonance.
   The dashed lines show the frequencies $\omega_{\rm pl}^0 ({\bf q}) = \sqrt{(2 \pi e^2 q/m)n_{\alpha \alpha}^{\rm intra}}$,
   with $V_0 n_{\alpha \alpha}^{\rm intra} =0.109$ (inset)
   and $\omega_{\rm pl}^{\rm tot,0} ({\bf q}) = \sqrt{(2 \pi e^2 q/m)n_{\alpha \alpha}^{\rm tot}}$,
   with $V_0 n_{\alpha \alpha}^{\rm tot}  = 1.045$ (main figure).
   The parameters are the same as in Fig.~9.
 }
  \end{figure}

However, the best way to study the finite ${\bf q}$ effects in the dielectric function 
{{{\rm on an equal footing with}}} 
the dc and dynamical ${\bf q} \approx 0$ conductivity,
is to use the usual Fermi liquid expression for $\sigma_{\alpha \alpha}^{\rm intra} ({\bf q},\omega)$
[given by Eq.~(47) from Ref.~\cite{Kupcic14}, with $M^{LL}({\bf q}, {\bf k}, \omega) \approx {\it i} \Gamma_1$, for example]
and the current-dipole expression for $\sigma_{\alpha \alpha}^{\rm inter} ({\bf q},\omega)$.
	Figure 10 shows the two-dimensional plot of the energy loss function 
\begin{eqnarray}
&& \hspace{-5mm}
-{\rm Im} \bigg\{ \frac{1}{\varepsilon ({\bf q}, \omega)} \bigg\} =
\frac{{\rm Im} \{\varepsilon ({\bf q}, \omega)\}}{|\varepsilon ({\bf q}, \omega)|^2}, 
\label{eq37}
\end{eqnarray}
for $\varepsilon ({\bf q}, \omega)$ obtained in the described way.
	The result 
{{{\rm is typical of}}}
two-dimensional multiband electronic systems with wide bands.
	The intraband plasmon mode is clearly visible in the ${\bf q}, \omega$ region in which the Landau damping is absent.
	{{{\rm For long wavelengths,}}} 
the frequency $\omega_{\rm pl} ({\bf q})$ is close to the bare intraband plasmon frequency $\omega_{\rm pl}^0 ({\bf q})$, 
because the dynamical screening effects of the rest of the $\pi$ electrons is negligible in this case \cite{Despoja13,Kupcic14}.

The interband plasmon resonance in the energy loss function (\ref{eq37})
in two-dimensional two-band systems exists only for large enough wave vectors ($q > q_1$).
	Since the interband plasmon frequency $\omega_{\rm pl}^{\rm tot} ({\bf q})$ 
is the second root of the real part of $\varepsilon ({\bf q}, \omega)$, it is expected to be clearly visible in
${\rm Re} \{\varepsilon ({\bf q}, \omega)\}$, at least in the ideal conductivity limit and for large enough wave vectors
(see Fig.~4 in Ref.~\cite{Novko15}).
	{{{\rm As may be anticipated from}}}
the definition relation 
\begin{eqnarray}
&& \hspace{-5mm}
[\omega_{\rm pl}^{\rm tot}({\bf q})]^2 =
4q \int_0^{\omega_{\rm pl}^{\rm tot}}
d \omega' \, \frac{[\omega_{\rm pl}^{\rm tot}({\bf q})]^2}{[\omega_{\rm pl}^{\rm tot}({\bf q})]^2 - \omega'^2}
{\rm Re} \{ \sigma_{\alpha \alpha} ({\bf q}, \omega') \}
\nonumber \\
&& \hspace{3mm}
+ 4q \int_{\omega_{\rm pl}^{\rm tot}}^{\infty} d \omega' \, 
\frac{[\omega_{\rm pl}^{\rm tot}({\bf q})]^2}{[\omega_{\rm pl}^{\rm tot}({\bf q})]^2 - \omega'^2}
{\rm Re} \{ \sigma_{\alpha \alpha} ({\bf q}, \omega') \},
\label{eq38}
\end{eqnarray}
{{{\rm there are two different regimes, depending upon whether}}}
$[\omega_{\rm pl}^{\rm tot}({\bf q})]^2/[[\omega_{\rm pl}^{\rm tot}({\bf q})]^2 - \omega'^2]$ in the first term in Eq.~(\ref{eq38}) 
is equal to unity or not \cite{Landau95,KupcicEK}.
	In the first regime  ($q_2 < q$) we have 
$\omega_{\rm pl}^{\rm tot} ({\bf q}) \approx \omega_{\rm pl}^{\rm tot,0} ({\bf q})
= \sqrt{(2 \pi e^2 q/m)n_{\alpha \alpha}^{\rm tot}({\bf q})}$,
while in the second regime ($q_1 < q < q_2$) the frequency $\omega_{\rm pl}^{\rm tot} ({\bf q})$ 
is well above the bare interband plasmon frequency $\omega_{\rm pl}^{\rm tot,0} ({\bf q})$.
	According to Fig.~10, in graphene the wave vector $q_1$ is approximately 
equal to $0.05/a_0$ and the wave vector $q_2$ is well above $0.2/a_0$.
	For $q < q_1$, the collective peak in $-{\rm Im} \{ 1/\varepsilon ({\bf q}, \omega) \}$ associated with the interband plasmon resonance 
transforms into the Van Hove single-particle peak in ${\rm Im} \{ \varepsilon ({\bf q}, \omega) \}$ 
(placed at $\hbar \omega \approx 2t$ \cite{Stauber10,Kupcic14}).
	The solid line in the main figure shows the position of such a composite interband resonance from $q=0$ up to $q = 0.2/a_0$.	
	This change of character of the interband $\pi$ excitations in the energy loss function 
was studied in pristine graphene in Ref.~\cite{Novko15} within the common charge-charge approach.
	The present study of doped graphene gives qualitatively the same result:
the $q^2$ dependence of the single-particle peak at $q \approx 0$ and the $\sqrt{q}$ dependence 
of the collective resonance at large enough wave vectors.
	Here we show how the prefactor in the $\sqrt{q}$ regime is related to the partial transverse conductivity sum rule.

\section{Conclusion}
In this paper, we have shown that it is possible to simplify the analysis of electrodynamic properties 
of pristine and doped graphene by using the quantum transport equations 
for auxiliary electron-hole propagators \cite{Kupcic13,Vollhardt80}
instead of the original Bethe-Salpeter equations.
{{{\rm The key to better understanding of}}}
electrodynamic properties of graphene is to solve the quantum transport equations 
{{{\rm in a way consistent with}}}
the charge continuity equation.
	In such an approach, the Ward identity relations 
{{{\rm play an essential role in determining}}}
the exact form of the total number of charge carriers in the partial transverse conductivity sum rule.
	As in any multiband case, this effective number consists of the intraband and interband contributions.
	However, in graphene, as well as in similar multiband electronic systems in which the threshold energy 
for interband electron-hole excitations is negligible, these two contributions are equally important 
when analyzing the dc conductivity and the intraband and interband plasmon resonances.
	They have a structure which is different from the nominal concentration of conduction electrons 
in Fermi liquid theory, but their role in describing transport coefficients 
and the dynamical conductivity is very similar.

We also shown that the current-dipole conductivity formula, which is intimately related with these 
quantum transport equations, represents the most natural way to take into account the effects 
of vertex corrections.
	In principle, this can be done by using the relaxation-time approximation, 
not only in clean but also in dirty systems.
	We demonstrate the advantages of using the current-dipole conductivity formula over other methods
(the widely used current-current approach, for example) 
by considering several open questions regarding electrodynamic 
properties of pristine and doped graphene:
the dc conductivity of ultraclean and dirty lightly doped samples \cite{Bolotin08}, 
the dynamical conductivity of moderately doped samples \cite{Li08}, and the dispersions of Dirac and $\pi$ plasmon resonances 
in both pristine and doped samples \cite{Yan13,Liou15}.

\section*{Acknowledgments}
The authors thank D. Novko for technical support.
This research was supported by the Croatian Ministry of Science, Education and Sports
under Project No. 119-1191458-0512 and the University of Zagreb grant No. 202758.

\appendix
\section{Current-current approach without vertex corrections}
After neglecting the current vertex renormalizations in the Bethe-Salpeter equations (\ref{eq1}), 
the current-current contribution to the conductivity 
tensor (\ref{eq13}) can be represented by the first diagram in the second row of Fig.~2 and written in the form
\begin{eqnarray}
&& \hspace{-5mm} 
\Delta \sigma_{\alpha \alpha}^0 ({\bf q}, {\it i}\nu_n)  = \frac{\it i}{\omega}
\sum_{LL'} \frac{1}{V} \sum_{{\bf k} \sigma}
|J^{LL'}_{\alpha}({\bf k},{\bf k}_+)|^2 {\cal G}_L ({\bf k}, {\it i}\omega_n)
\nonumber \\
&& \hspace{20mm}
\times  {\cal G}_{L'} ({\bf k}_+, {\it i}\omega_{n+}).
\label{eqA1}
\end{eqnarray}
	This conductivity formula 
{{{\rm represents widely applicable model for analyzing}}}
electrodynamic properties of doped graphene \cite{Ando98,Ando02,Peres08,Carbotte10}.
	It depends on ${\cal A}_L ({\bf k}, \varepsilon)$ directly, and not through the momentum distribution function
$n_L ({\bf k})$.
	The main disadvantage of this approach is that it is focused only on the indirect contributions to 
$\sigma_{\alpha \alpha}^{\rm intra} ({\bf q}, {\it i}\nu_n)$, and, consequently,
{{{\rm does not apply to}}}
finite wave vectors ${\bf q}$.

The real part of the analytically continued form of Eq.~(\ref{eqA1}) at ${\bf q} = {\bf 0}$ 
can be represented by the following textbook expression \cite{Mahan90,Ando98}
\begin{eqnarray}
&& \hspace{-10mm} 
{\rm Re} \{ \sigma_{\alpha \alpha}^0 (\omega) \} =
\int_{-\infty}^\infty \frac{ d \varepsilon}{4 \pi}
\sum_{LL'} \frac{1}{V} \sum_{{\bf k} \sigma} \hbar |J^{LL'}_{\alpha}({\bf k},{\bf k}_+)|^2  
\nonumber \\
&& \hspace{2mm}
\times {\cal A}_{L} ({\bf k}, \varepsilon) {\cal A}_{L'} ({\bf k}_+, \varepsilon_+)
\frac{f(\varepsilon) - f(\varepsilon + \hbar \omega)}{\hbar \omega}
\nonumber \\
&& \hspace{0mm}
\equiv \int_{-\infty}^\infty \frac{ d \varepsilon}{4 \pi}
\frac{f(\varepsilon) - f(\varepsilon + \hbar \omega)}{\hbar \omega}
{\cal P}_{\alpha \alpha} (\varepsilon, \varepsilon + \hbar \omega).
\label{eqA2}
\end{eqnarray}
	Here,
\begin{eqnarray}
&& \hspace{-10mm} 
{\cal P}_{\alpha \alpha} (\varepsilon, \varepsilon + \hbar \omega) 
\approx {\cal P}_{\alpha \alpha} ({\bf q} \approx 0, \varepsilon, \varepsilon + \hbar \omega)
\label{eqA3}
\end{eqnarray}
is the auxiliary $T=0$ current-current correlation function, with 
\begin{eqnarray}
&& \hspace{-10mm} 
{\cal P}_{\alpha \alpha} ({\bf q},\varepsilon, \varepsilon + \hbar \omega) = 
\sum_{LL'} \frac{1}{V} \sum_{{\bf k} \sigma} \hbar |J^{LL'}_{\alpha}({\bf k},{\bf k}_+)|^2  
\nonumber \\
&& \hspace{20mm}
\times {\cal A}_{L} ({\bf k}, \varepsilon) {\cal A}_{L'} ({\bf k}_+, \varepsilon_+)
\label{eqA4}
\end{eqnarray}
and
$\varepsilon_+ = \varepsilon + \hbar \omega$.

 \begin{figure}
   \centerline{\includegraphics[width=17pc]{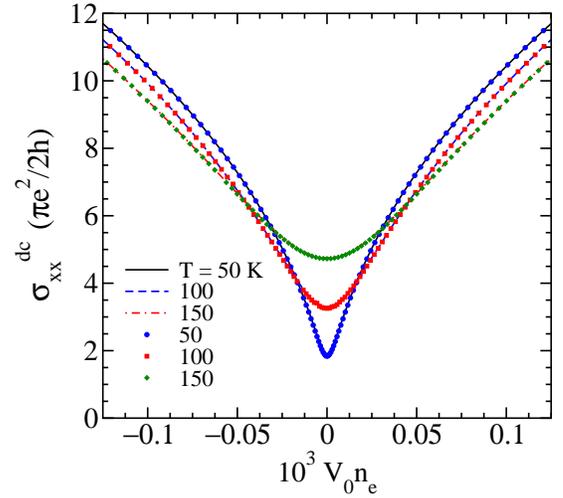}}
   \caption{The doping dependence of $\sigma_{\alpha \alpha}^{\rm dc}$ at $T = 50$, 100, and 150 K.
   The solid, dashed, and dot-dashed lines represent the results of the current-dipole conductivity formula (\ref{eq25}) for
   $\hbar \Gamma_1 = \hbar \Gamma_2 = 5$ meV and $n_L ({\bf k}) = f_L ({\bf k})$.
   The full circles, squares, and diamonds are the results of the current-current conductivity formula (\ref{eqA2})
   for $\hbar \Sigma^i = 2.5$ meV.
}
  \end{figure}

The full circles, squares, and diamonds in Fig.~11 show the dc conductivity in graphene calculated in the Dirac cone approximation by using
Eq.~(\ref{eqA2}), for $\hbar \Sigma^i = 2.5$ meV.
	This figure shows that the current-dipole conductivity formula (\ref{eq25}) and the current-current 
conductivity formula (\ref{eqA2}) give essentially the same results, when the damping energies 
in these expressions are mutually related by $\hbar \Gamma = 2 \hbar \Sigma^i$, with $n_L({\bf k}) = f_L({\bf k})$,
and when the temperature is not too low.
	Namely, it is well known that Eq.~(\ref{eqA2}) is characterized by the ballistic conductivity $(8/\pi^2) (\pi e^2/ 2 h)$ 
at $E_{\rm F} = 0$ and $T=0$ \cite{Ando98},
which is in disagreement with the ballistic conductivity of the current-dipole conductivity formula (\ref{eq25}), $(\pi e^2/ 2 h)$ \cite{Ziegler06}, 
as well as with experiment, $(8/\pi) (\pi e^2/ 2 h)$ \cite{Novoselov05}.
	Another important difference between these two conductivity formulas is in the structure of the $T=0$ dc conductivity:
$\sigma^{\rm dc}_{\alpha \alpha} = \sigma^{\rm dc, inter}_{\alpha \alpha}$ in the current-dipole approach,
and $\sigma^{\rm dc}_{\alpha \alpha} = 2 \sigma^{\rm dc, intra}_{\alpha \alpha} = 2 \sigma^{\rm dc, inter}_{\alpha \alpha}$
in the current-current approach.

\section{Phenomenological treatment of vertex effects}
We can use the identity relation
\begin{eqnarray}
&& \hspace{-5mm}
{\cal G}_{L} ({\bf k}, \varepsilon+ s {\it i} \eta) {\cal G}_{L'} ({\bf k}_+, \varepsilon_+  + s' {\it i} \eta) 
\nonumber \\
&& \hspace{0mm}
=\frac{{\cal G}_{L} ({\bf k}, \varepsilon+ s {\it i} \eta) - {\cal G}_{L'} ({\bf k}_+, \varepsilon_+ + s' {\it i} \eta)}{
\omega + \varepsilon_{LL'}^0 ({\bf k},{\bf k}_+)/\hbar + s \Sigma_L ({\bf k},\varepsilon) 
- s' \Sigma_{L'} ({\bf k}_+,\varepsilon_+)}
\nonumber \\
\label{eqB1}
\end{eqnarray}
to obtain
\begin{eqnarray}
&& \hspace{-5mm} 
{\cal P}_{\alpha \alpha} ({\bf q},\varepsilon, \varepsilon + \hbar \omega) = 
- \sum_{ss'} s s' \sum_{LL'} \frac{1}{V} \sum_{{\bf k} \sigma}
|J^{LL'}_{\alpha}({\bf k},{\bf k}_+)|^2  
\nonumber \\
&& \hspace{5mm}
\times \frac{{\cal G}_{L} ({\bf k}, \varepsilon+ s {\it i} \eta) - {\cal G}_{L'} ({\bf k}_+, \varepsilon_+ + s' {\it i} \eta)}{
\hbar \omega + \varepsilon_{LL'}^0 ({\bf k},{\bf k}_+) + s \hbar \Sigma_L ({\bf k},\varepsilon) 
- s' \hbar \Sigma_{L'} ({\bf k}_+,\varepsilon_+)}.
\nonumber \\
\label{eqB2}
\end{eqnarray}
	Here, $\Sigma_L ({\bf k},\varepsilon)$ is the single-electron self-energy from Eq.~(\ref{eq19}), which is the solution of the 
corresponding Dyson equation for ${\cal G}_{L} ({\bf k}, \varepsilon+  {\it i} \eta)$ from Fig.~1(b).

The resulting expression for ${\cal P}_{\alpha \alpha} ({\bf q}, \varepsilon, \varepsilon_+)$ is
\begin{eqnarray}
&& \hspace{-10mm} 
{\cal P}_{\alpha \alpha} ({\bf q},\varepsilon, \varepsilon + \hbar \omega) \approx 
- \sum_{s s'}  s s' \sum_{LL'} \frac{1}{V} \sum_{{\bf k} \sigma}
|J^{LL'}_{\alpha}({\bf k},{\bf k}_+)|^2  
\nonumber \\
&& \hspace{10mm}
\times \frac{{\cal G}_{L} ({\bf k}, \varepsilon+ s {\it i} \eta) - {\cal G}_{L'} ({\bf k}_+, \varepsilon_+ + s' {\it i} \eta)}{
\hbar \omega + \varepsilon_{LL'} ({\bf k},{\bf k}_+) + {\it i} \hbar \Gamma^{LL'}_{ss'} ({\bf k})}, 
\label{eqB3}
\end{eqnarray}
with $\Gamma^{LL'}_{ss'} ({\bf k}) = s \Sigma_L^i ({\bf k}) - s' \Sigma_{L'}^i ({\bf k})$.
	The expression (\ref{eqA2}), together with Eqs.~(\ref{eqA3}) and (\ref{eqB3}), is the forth important result of the present analysis.
	It represents the usual current-current formula for the conductivity tensor, which is shown in the form
directly related to the current-dipole conductivity formula (\ref{eq25}).
	Both of these formulas contain two damping energies:
$\Sigma_L^i ({\bf k})$ in the spectral functions in the numerator, and $\Gamma^{LL'}({\bf k})$ in the denominator.
	Evidently both of them 
{{{\rm are first order in}}} 
the spectral functions ${\cal A}_L ({\bf k}, \varepsilon)$.

\end{document}